\documentclass[a4paper,fleqn,usenatbib]{mnras}
 
\usepackage{newtxtext,newtxmath}
 
\usepackage[T1]{fontenc}
\usepackage{ae,aecompl}

\usepackage{graphicx}   
\usepackage{amsmath}    
\usepackage{amssymb}    
 
\title[System parameters of Sco X-1]{Precision ephemerides for gravitational-wave searches -- III. Revised system parameters of Sco X-1}

\author[L. Wang et al.]{
L. Wang,$^{1}$\thanks{E-mail: \href{mailto:Zhuqing.Wang@warwick.ac.uk}{Zhuqing.Wang@warwick.ac.uk} (LW)}
D. Steeghs,$^{1,2}$
D. K. Galloway,$^{2,3}$
T. Marsh$^{1}$ and
J. Casares$^{4,5}$
\\
$^{1}$Department of Physics, University of Warwick, Gibbet Hill Road, Coventry CV4 7AL, UK\\
$^{2}$School of Physics \& Astronomy, Monash University, VIC 3800, Australia\\
$^{3}$Monash Centre for Astrophysics, Monash University, VIC 3800, Australia\\
$^{4}$ Instituto de Astrof\'isica de Canarias, 38205 La Laguna, Tenerife, Spain\\
$^{5}$ Departamento de astrof\'isica, Univ. de La Laguna, E-38206 La Laguna, Tenerife, Spain
}

\date{Accepted XXX. Received YYY; in original form ZZZ}
 
\pubyear{2015}
 
\begin{document}
\label{firstpage}
\pagerange{\pageref{firstpage}--\pageref{lastpage}}
\maketitle
 
\begin{abstract}
Neutron stars in low-mass X-ray binaries are considered promising candidate sources of continuous gravitational-waves.
These neutron stars are typically rotating many hundreds of times a second. The process of accretion can potentially generate and support non-axisymmetric distortions to the compact object, resulting in persistent emission of gravitational-waves. 
We present a study of existing optical spectroscopic data for Sco X-1, a prime target for continuous gravitational-wave searches, 
with the aim of providing revised constraints on key orbital parameters required for a directed search with advanced-LIGO data.
From a circular orbit fit to an improved radial velocity curve of the Bowen emission components, we derived an updated orbital period and ephemeris. 
Centre of symmetry measurements from the Bowen Doppler tomogram yield a centre of the disc component of 90 km $\mathrm{s^{-1}}$, which we interpret as a revised upper limit to the projected orbital velocity of the NS $K_1$.
By implementing Monte Carlo binary parameter calculations, and imposing new limits on $K_1$ and the rotational broadening, we obtained a complete set of dynamical system parameter constraints including a new range for $K_1$ of 40--90 km $\mathrm{s^{-1}}$.
Finally, we discussed the implications of the updated orbital parameters for future continuous-waves searches.
\end{abstract}
 
 
 
 
\section{Introduction}
 
Low-mass X-ray binaries (LMXBs) are interacting binary systems comprising a low-mass donor star transferring matter on to a compact object (either a neutron star or a black hole) via an accretion disc.
They can be divided into two populations based on their long-term behavior; those that accrete \emph{persistently} at high mass-accretion rates ($L_X$ $\sim$ $10^{36}$ erg $\mathrm{s^{-1}}$), 
and the \emph{transient} systems which spend most of their lives in a dim quiescent state, but undergo sporadic outbursts when their X-ray luminosity increases. 
Neutron stars (NSs) in binaries are considered promising sources of gravitational-wave (GW) emission for ground-based GW detectors.
Not only are these NSs typically rotating many hundreds of times a second,
the process of accretion can potentially generate and support non-axisymmetric distortions to the compact object (e.g. \citealt{1998ApJ...501L..89B}).
A quadrupole mass moment will lead to persistent emission of GWs at twice the NS spin frequency.

Sco X-1 was
the first LMXB discovered outside the Solar System \citep{1962PhRvL...9..439G} and the brightest persistent X-ray source in the sky.
The orbital period of 18.9 hr was first reported by \citet*{1975ApJ...195L..33G}
from photometry of the optical counterpart V818 Sco (V $\sim$ 12.5) using 1068 plates spanning over 84 yr.
\citet*{1999ApJ...512L.121B} measured the trigonometric parallax from Very Long Baseline Array (VLBA) radio observations, and obtained a distance of 2.8 $\pm$ 0.3 kpc.
Further observations revealed the presence of twin radio lobes, which led to an estimate for the inclination of $44^{\circ}$ $\pm$ $6^{\circ}$, assuming that the radio jet is perpendicular to the orbital plane \citep*{2001ApJ...558..283F}.
It is believed to harbor a NS based on its X-ray behavior.
However, no coherent pulsations or bursts have been observed so far from this source. 

Sco X-1 has been a prime target for continuous GW searches for the Laser Interferometer Gravitational-Wave Observatory (LIGO).
Due to its relative proximity to Earth and high mass accretion rate, it is predicted to be the strongest GW emitter among the family of known LMXBs within the relevant frequency passband \citep{2007PhRvD..76h2001A}.
A key difficulty in searching for the GWs emitted by NSs in LMXBs is the lack of precise knowledge about the position and velocity of the NS in the binary orbit \citep{2008MNRAS.389..839W}.
Searching over possible values of unknown parameters carries a cost in a computationally-bound search, and many methods carry a trade-off between computational cost and sensitivity. 
The smaller the parameter space region that needs to be searched, the more sensitive the search can be made, given fixed computing resources. 
Without accurate determinations of the orbital parameters, a $\emph{directed}$ search may still be carried out, but an observational `penalty' must be paid.
Effectively, the signal must be proportionately stronger, compared to a source where the parameters are more precisely constrained, in order to reach the same order of confidence for a detection.

The most important orbital parameters required by a directed search include the orbital period ($\mathrm{P_{orb}}$), the absolute phase of the system (for which we quote the epoch of inferior conjunction of the companion star $T_0$), and the eccentricity $e$ (e.g., \citealt{2008MNRAS.389..839W}). The projected semi-major axis of the NS ($a_x$sin$i$) must also be searched over (\citealt{2015PhRvD..92b3006M}; \citealt{2015PhRvD..91j2003L}; \citealt{2015PhRvD..91j2005W}; \citealt{2016CQGra..33j5017M}; \citealt{2014PhRvD..89d3001S}), and is related to the amplitude of the projected orbital velocity of the NS, $K_1$ (or $v_x$sin$i$), as $K_1$ = 2$\pi$ $a_x$sin$i$/$\mathrm{P_{orb}}$.
For systems in active states, the optical emission is dominated by the reprocessing of hard X-rays in the outer accretion disc, which severely hampers the detection of the much fainter companion.
A recent study by \citet{2015MNRAS.449L...1M} found no evidence of companion spectral features for Sco X-1 even in the near-infrared (where the companion star could make a substantial contribution).
A novel avenue for obtaining dynamical information on Sco X-1 was opened up thanks to the discovery of extremely narrow, high-excitation emission lines arising from the X-ray illuminated atmosphere of the donor star (\citealt{2002ApJ...568..273S}, hereafter SC02).
These narrow components were strongest in the Bowen region (consisting of a blend of \ion{N}{iii}/\ion{C}{iii} lines), 
and are the result of fluorescence of the gas by UV photons from the hot inner disc \citep*{1975ApJ...198..641M}.
Since the first detection of the donor star of Sco X-1 in emission, it has been shown for several other persistent systems that the use of Bowen fluorescence lines as tracers of counterpart radial velocities provides a unique opportunity to pursue robust radial velocity studies (e.g. \citealt{2006MNRAS.373.1235C}; \citealt{2007MNRAS.380.1182B}).

Our project -- Precision Ephemerides for Gravitational-wave Searches (PEGS) -- aims at 
providing the most reliable constraints for the key orbital parameters for candidate persistent GW sources.   
\citet{2014ApJ...781...14G} (hereafter G14) presented the results of a pilot program for improving the precision of the orbital period and ephemeris of the most promising system, Sco X-1, using the proven Bowen diagnostic of the mass donor.
In this follow-up study to G14, we provide revised constraints on key orbital parameters (derived from a re-analysis of existing optical spectroscopic data) in direct support of GW observations of Sco X-1 in the Advanced-LIGO era.
We also describe new analysis tools that are relevant to the use of emission lines as dynamical probes of the donor stars in compact binaries in general.

\section{Source data and RV determination}
\label{sec:sourcedata}
 
This study makes use of all available spectroscopic data previously obtained for Sco X-1 and initially presented in SC02 and G14.
Two epochs of data separated by 12 years (1999 -- 2011) were obtained using the Intermediate dispersion Spectrograph and Imaging System (ISIS) at the 4.2m William Herschel Telescope (WHT).
In addition, Sco X-1 was observed at 44 random epochs between 2011 May 29 -- August 23 with the Ultraviolet and Visual Echelle Spectrograph (UVES) on the European Southern Observatory (ESO) Very Large Telescope (VLT; programme 087.D-0278).
The superior quality of the combined dataset allowed the application of several techniques for obtaining binary parameter constraints, enabling us to go beyond the initial results provided in G14.
Table~\ref{tab:logobs} shows a summary of source data.
For details of the observations and data reduction procedures, we refer to SC02 and G14.
 
We first reanalyzed the Bowen emission components from the irradiated front face of the donor using the well-established radial velocity (RV) fitting method.
Following SC02 and G14, we used the $\textsc{molly}$\footnote{\url{http://deneb.astro.warwick.ac.uk/phsaap/software}} $\textit{mgfit}$ routine for fitting multiple Gaussians to time-resolved, continuum-normalised spectra
and (for each Bowen profile) measuring the common RV of the three strongest narrow lines (\ion{N}{iii} $\lambda$$\lambda$4634 \& 4640$\AA$ and \ion{C}{iii} $\lambda$4647).
For the WHT data, we used the same multi-Gaussian model (with 6 free parameters) as was used in SC02 and G14.
It is known that during certain phase ranges, especially near phase zero, the narrow lines could be much weaker than at other phases. The weakness of the narrow lines had such a dramatic impact on the RV fitting in G14 (see G14; Fig. 5) that
the majority of the WHT measurements within $\pm$ 0.1 of phase 0 (i.e. the inferior conjunction of the donor star) were considered spurious and were thus eliminated from further analyses.
To salvage some of the data points in this range, we can go one step further and take into account the spectrum-to-spectrum variation.
For each spectrum, the expected RV of the narrow lines was calculated (using a sinusoidal model) and used as the initial fit value of RV to guide the optimizer to a reasonable optimum.
The quality of individual Gaussian fits was also carefully assessed by visual inspection, especially for spectra taken near the inferior conjunction of the companion.
 
In order to exploit the potential of our highest resolution VLT data (see, e.g., G14, Fig. 4), 
we modified the model
for extracting UVES velocities.
Instead of fixing the FWHM of all the Gaussians, we fixed the width of the underlying broad component to the value used in G14 but left the (common) width of the narrow components free to vary.
This yielded a revised set of UVES RVs that are fully consistent with the values reported in G14 (within the formal errors) as well as the fitted width of the sharp \ion{N}{iii}/\ion{C}{iii} components.
We found that the model with 7 free parameters fitted the UVES spectra better, and that the average statistical uncertainty of the VLT measurements was reduced to 1.1 km $\mathrm{s^{-1}}$ (compared to the previously reported precision of 2.2 km $\mathrm{s^{-1}}$).
       
\begin{table*}
        \centering
        \caption{Observations of Sco X-1.}
        \label{tab:logobs}
        \begin{tabular}{l l c c c c c}
                \hline
                \hline
                {Date} & {Instrument} & {Exposure time (s)} & {$N_{\mathrm{obs}}$} & {Phase coverage} & {Resolution (\AA)} & {Reference} \\
                \hline
                1999 Jun 28--30     & ISIS/WHT & 300 & 137 & 75\% & 0.84 & SC02 \\
                2011 Jun 16--18     & ISIS/WHT & 300 & 157 & 75\% & 0.32 & G14 \\
                2011 May 29--Aug 23 & UVES/VLT & 720 & 44  & 34\% & 0.1  & G14 \\
                \hline
        \end{tabular}
\end{table*}


 

\section{Determination of observables}
\label{sec:analysis} 
The number of unreliable RVs that deviate significantly from the predicted values after applying the modified approach is substantially lower than previously estimated.
Narrow lines from the donor are still present and can be detectable within the phase ranges $-$0.1 to $+$0.1 (albeit being rather weak since we are facing the un-irradiated side of the donor).
The derived RVs within those ranges have noticeably larger error bars than data points at other phases but ought to be included in the RV fit (with correspondingly smaller weights).
For these reasons, we refrain from screening data (as was introduced in G14 to eliminate a large fraction of spurious RVs), and choose to keep the total of 338 WHT \& VLT measurements in the RV analyses of this paper to derive a revised set of Sco X-1 system parameters.
The revised and refined RV datasets are provided in the supporting information available online.
 
\subsection{Circular orbit fit}
\label{sec:sinfit} 
 
Assuming a negligible (intrinsic) eccentricity of the binary orbit ($e$ = 0), the simplest model for the variation of the projected Bowen RVs is a sinusoid of the form
\begin{equation}
V(t) = K\, \mathrm{sin}\bigg(\frac{2 \pi (t - \mathrm{T_0})}{\mathrm{P_{orb}}}\bigg) + \gamma,
        \label{eq:RVsin}
\end{equation}
where K is the velocity semi-amplitude, $\mathrm{T_0}$ is the epoch of inferior conjunction of the mass donor, $\mathrm{P_{orb}}$ is the orbital period, and $\gamma$ is the RV of the system (systemic velocity).

To fit the 4 free parameters ($\mathrm{T_0}$, $\mathrm{P_{orb}}$, K, and $\gamma$), we started with a least-squares method where each data point was weighted according to the associated error bar given by $\textit{mgfit}$ whilst giving no differentiation between separate VLT \& WHT datasets.
The reduced $\chi^2$ value we obtained ($\chi^2_{\nu}$ = 11.3 for 334 degrees of freedom) was greater than that obtained in G14 ($\chi^2_{\nu}$ = 8.5 for 238 degrees of freedom) due to the inclusion of all the RV measurements, and much greater than unity.
A high value of the $\chi^2$-statistic can either be caused by limitation of the sine-wave model, or reflect the fact that the initial measurement uncertainties on the RVs are underestimated.
The fit residuals we see are not consistent with those expected from irradiation effects (see Section~\ref{sec:discuss}). Therefore, for the purpose of parameter estimation, we considered that the measurement errors had been systematically underestimated by some fractional amount (efac).
 
Hence we proceeded to introduce additional `multiplicative factors' into the sinusoidal model for adjusting errors on the RV data.
To recognize that the data consists of three datasets with different spectral resolution/SNR contributions (1999 WHT, 2011 WHT and 2011 VLT; see Table~\ref{tab:logobs}), we invoked 3 multiplicative factors in total to rescale errors on individual multi-Gaussian fits per dataset.
The 1999 and 2011 WHT observations (both covering 75\% of the orbital cycle) were then refitted \textit{separately} with the Bayesian Markov Chain Monte Carlo (MCMC) approach (using the \textsc{python} package \texttt{emcee}; \citealt{2013PASP..125..306F}).
In both cases, the resulting posterior distribution of the multiplicative factor was Gaussian, therefore we took the posterior mean ($\mathrm{efac_{1999}}$ = 4.3 and $\mathrm{efac_{2011}}$ = 2.5) as the best estimate of efac for the WHT datasets.
It is also worth noting that the RV semi-amplitude for the 1999 data (77.6 $\pm$ 1.3 km $\mathrm{s^{-1}}$; 68 per cent) is significantly higher than the 2011 $\mathrm{K_{em}}$ velocity (72.0 $\pm$ 1.0 km $\mathrm{s^{-1}}$; 68 per cent).
This finding
is consistent with the discrepancy between the best-fit velocity semi-amplitude for the combined data ($\mathrm{K_{em}}$ = 74.9 $\pm$ 0.5; G14) and the result of SC02 ($\mathrm{K_{em,\,  1999}}$ = 77.2 $\pm$ 0.4 km $\mathrm{s^{-1}}$)
based on the 1999 observations alone;
and might be attributed to variations in the geometry of the line emission over epoch, as discussed in G14.
 
\begin{figure}
        \includegraphics[width=\columnwidth]{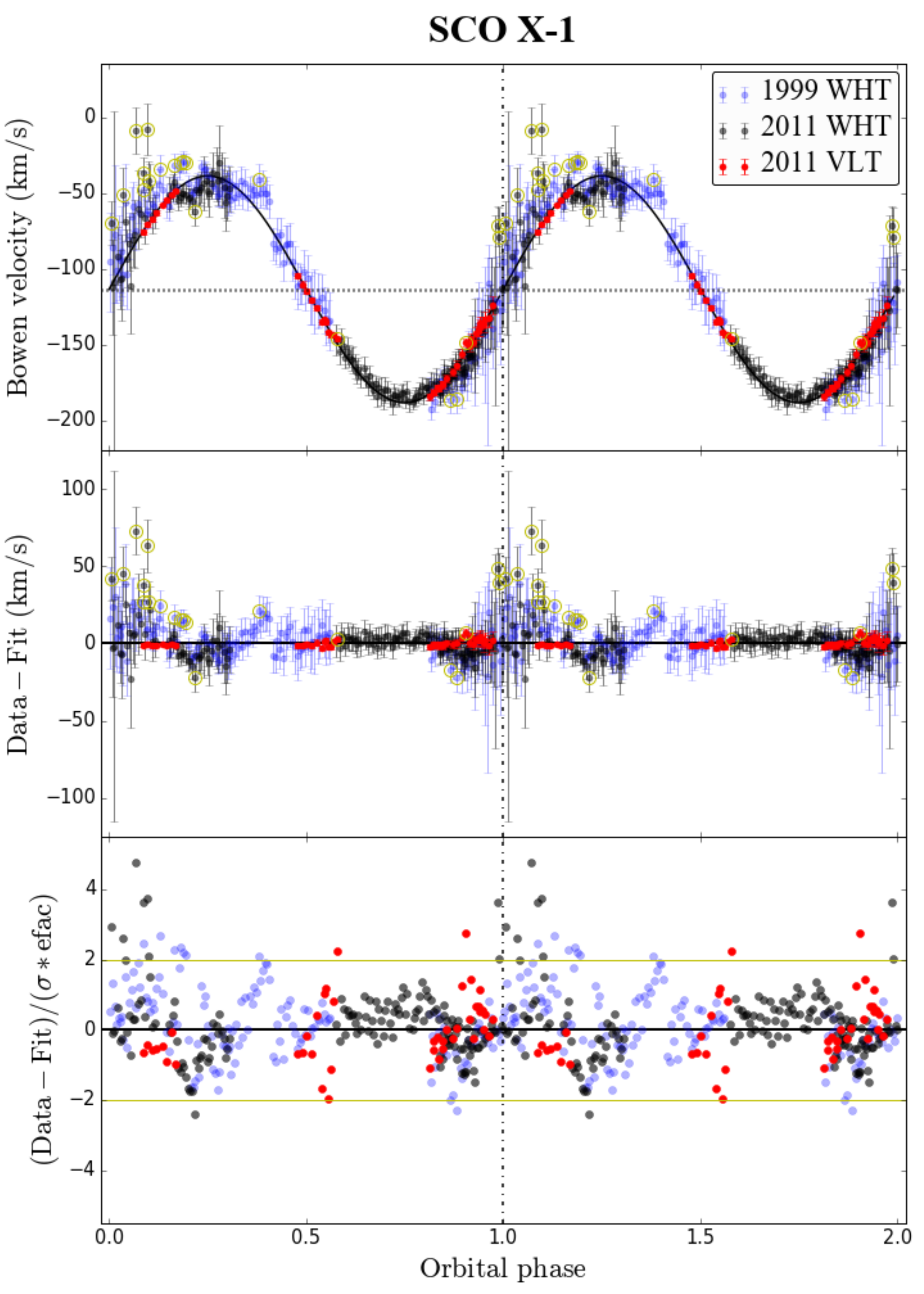}
    \caption{\emph{Top}: RVs of the Bowen emission components determined from the 1999 WHT (blue), 2011 WHT (black) and 2011 VLT (red) observations, with 1$\sigma$ error bars after scaling by efac. The best-fitting circular orbit model is overplotted in black. The systemic velocity is shown as a horizontal dotted line. Two cycles are shown for clarity purposes. Note that the RVs around phase zero have noticeably larger (rescaled) error bars (see Section~\ref{sec:sinfit}) than data points at other phases.  \emph{Middle}: the residuals as a function of the orbital phase. \emph{Bottom}: the residuals divided by rescaled error bars. Data points below/above the horizontal yellow line y = -2/y = 2 are identified as outliers and are marked by yellow circles in the upper and middle panels.}
    \label{fig:rvcurve}
\end{figure}

To determine conservative parameter estimates, we scaled the WHT errors (up) by the efac values derived from the MCMC fit and repeated the least-squares analysis of the combined WHT and VLT data.
At the final step, the VLT errors were scaled by a multiplicative factor ($\mathrm{efac_{VLT}}$ = 2.25). 
This ensures that the final reduced $\chi^2$ is equal to 1.0.
We also varied the initial parameter value of $\mathrm{T_0}$ to obtain the smallest possible cross-term in the covariance matrix between $\mathrm{P_{orb}}$ and $\mathrm{T_0}$ ($|$$V$($\mathrm{P_{orb}}$, $\mathrm{T_0}$)$|$ = 3 $\times$ $\mathrm{10^{-14}}$). 
This yielded revised 
orbital parameters consistent (to within $\lesssim$ 2.6$\sigma$) with those previously determined (by least-squares fitting to 242 RV measurements) in G14.
Fig.~\ref{fig:rvcurve} shows the improved RV curve 
(with the best fit sinusoid) 
consisting of a total of 338 measurements (cf. G14; Fig. 5).

The revised parameter estimates
are summarized as follows for direct comparison with G14:
\begin{center}
$\mathrm{T_0}$ = 2455522.6668 $\pm$ 0.0006 HJD(UTC) \\
$\mathrm{P_{orb}}$ = 0.7873132 $\pm$ 0.0000005 days \\
$\gamma$ = -113.6 $\pm$ 0.2 km $\mathrm{s^{-1}}$ \\
$\mathrm{K_{em,\, RV}}$ = 74.8 $\pm$ 0.4 km $\mathrm{s^{-1}}$. \\
\end{center}
To compare our new $\mathrm{T_0}$ constraint with that of G14, we added 1127 orbital cycles to give 2455522.6682 HJD (UTC), which differs from the estimate in this work by 121 seconds, within the 1$\sigma$ uncertainty of G14's estimate propagated to 2010.
The new $\mathrm{T_0}$ measurement improves the constraint on the absolute phase of the system by a factor of two (both by being more precise and by being less correlated with $\mathrm{P_{orb}}$ when considering times after 2010) relative to the result of G14. 
On the other hand, the $\mathrm{P_{orb}}$ measurement is at the same precision as that in G14, but differs from that value by 2.6$\sigma$. 
In Section~\ref{sec:bootstrapDT}, we perform tests to show that the new period and ephemeris constraints (derived from refined RV datasets) are indeed more accurate than previous findings. 

\subsection{Doppler tomography-based method}
\label{sec:MCD}
An alternative route towards orbital parameter estimation is provided by a technique known as Doppler tomography, developed by \citet{1988MNRAS.235..269M} for recovering accretion structures in mass-transferring binary systems.
The idea is that by observing time-dependent line profiles as a function of the binary orbit (preferably covering a complete orbital cycle), sufficient information can be obtained to invert the phase-resolved spectra into an `image' of emission distribution \emph{in Doppler/velocity coordinates}.
In a Doppler tomogram, the Roche-lobe-filling donor star is on the positive $V_y$-axis, shifted from the origin ($V_x$, $V_y$) = (0, 0) by its projected RV semi-amplitude ($K_2$);
the compact accretor (although often unseen) is below the origin at (0, -$K_1$);
the accretion disc still appears circular in velocity space, but is inside-out (i.e., the outer edge of the disc is nearer the origin).
Other major emission sites might be traced at an expected position, including the gas stream leaving the secondary and the stream-disc impact region.
For a review of the technique, see \citet{2001LNP...573....1M}; \citet{2004AN....325..185S}.
 
Since Doppler tomography uses all observed data simultaneously, it is ideally suited for weak features and provides the only way to detect the donor signature in the cases of low signal-to-noise ratios (e.g. \citealt{2008AIPC.1010..148C}; \citealt{2017MNRAS.466.2261W}).
The technique has also been applied to the high SNR data of Sco X-1 to reconstruct maps for a range of emission lines present in the optical spectrum (see SC02; Fig. 4).
The \ion{N}{iii} and \ion{C}{iii} maps in particular revealed a compact spot along the positive $V_y$-axis (consistent with an emission component from the irradiated donor),
from which an independent estimate of $\mathrm{K_{em}}$ could be accurately derived through a two-dimensional Gaussian fit.
 
\subsubsection{The Bowen map}
\begin{figure*}
        \includegraphics[width=2.0\columnwidth]{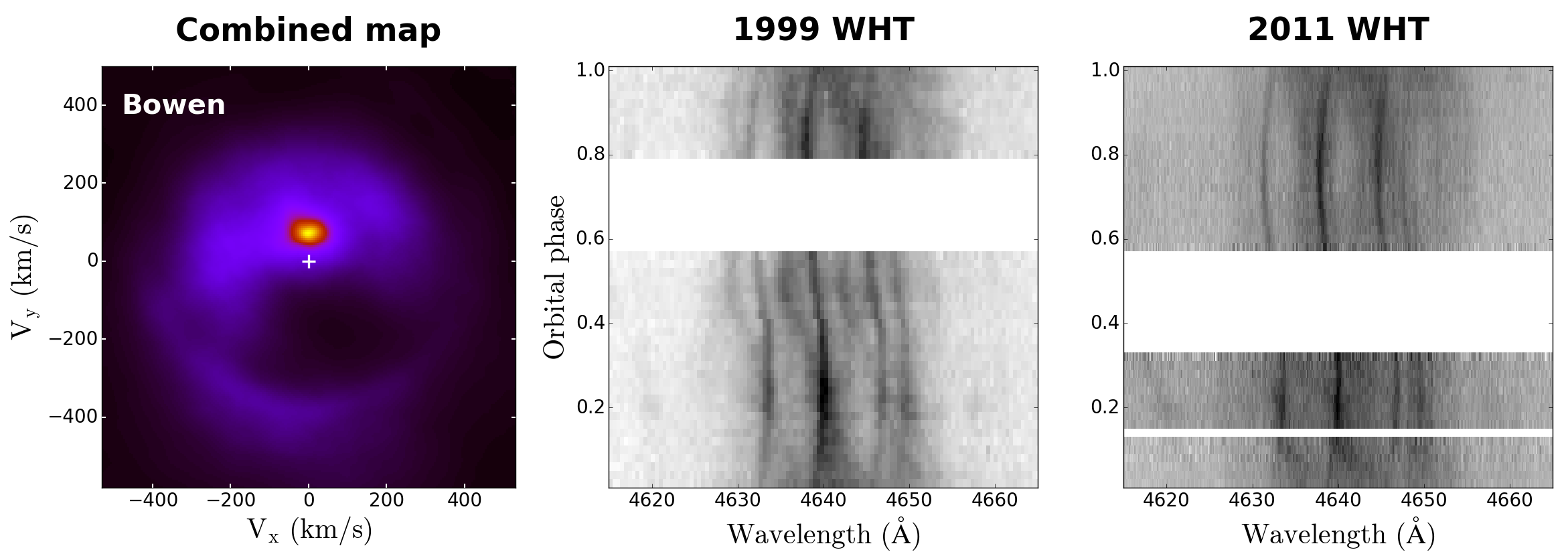}
    \caption{The combined Doppler tomogram of the Bowen blend (\emph{left}) for the 1999 \& 2011 WHT data (middle and right panels). The origin of the Doppler map (denoted by a white cross) corresponds to the center of mass of the system using $\gamma$ = -113.6 km $\mathrm{s^{-1}}$. The Bowen region is dominated by narrow `S-wave' components produced on the irradiated face of the companion.}
    \label{fig:dopmaps}
\end{figure*}

To enable a direct comparison to be made between the $\mathrm{K_{em}}$ amplitude obtained from the RV fitting and the tomography-based method, we constructed new \emph{Bowen Doppler images} (initially separately for the 1999 and 2011 WHT datasets; each covering 75\% of the orbital cycle) by assigning the three strongest Bowen emission lines to the same image, and updating the input parameters to Doppler tomography ($\gamma$, $\mathrm{T_0}$, and $\mathrm{P_{orb}}$) with the values determined in Section~\ref{sec:sinfit}.
Both 1999 and 2011 WHT Bowen maps showed a prominent spot at the expected location of the companion, and a much fainter ring-like feature, most likely originating from the accretion disc.
We note again that the centroid position of the 1999 donor emission spot is at a higher velocity ($V_x$ = 2 km $\mathrm{s^{-1}}$ and $V_y$ = 76 km $\mathrm{s^{-1}}$) than that inferred from the 2011 donor spot ($\mathrm{K_{em,\,  2011}}$ = 72.2 km $\mathrm{s^{-1}}$).
Furthermore, with the aid of the second-generation \textsc{doppler} code\footnote{\url{https://github.com/trmrsh/trm-doppler}} developed by T. Marsh (e.g., \citealt{2016MNRAS.455.4467M}), it is possible to construct one Bowen image for the \emph{combined} 1999 and 2011 WHT data even though these have different spectral resolutions.
The \emph{combined map} and the phase-binned, trailed spectrograms of the Bowen region (from the WHT observations) are displayed in Fig.~\ref{fig:dopmaps}.
 
\subsubsection{Bootstrap Doppler tomography}
\label{sec:bootstrapDT}
For robust estimation of $\mathrm{K_{em}}$ from the combined Bowen map we adopted the `bootstrap Monte Carlo' approach recently introduced by \citet{2017MNRAS.466.2261W} (hereafter, W17).
The method was initially developed for statistical uncertainty derivation and, most importantly, robust significance testing for spot features in the low SNR case of XTE J1814$-$338; but can be readily extended to the mid- to high-SNR regime.
Following the strategy described in W17, we created 2000 simulated images from 2000 independently generated bootstrapped datasets \citep{tEFR93a} in the same manner as for the original map (see W17, Section 5.2.2), using custom wrapper functions.
Measurements of various spot properties (e.g., the centroid position and the peak intensity) were then performed for the ensemble of combined images.
 
\begin{figure*}
        \includegraphics[width=2.0\columnwidth]{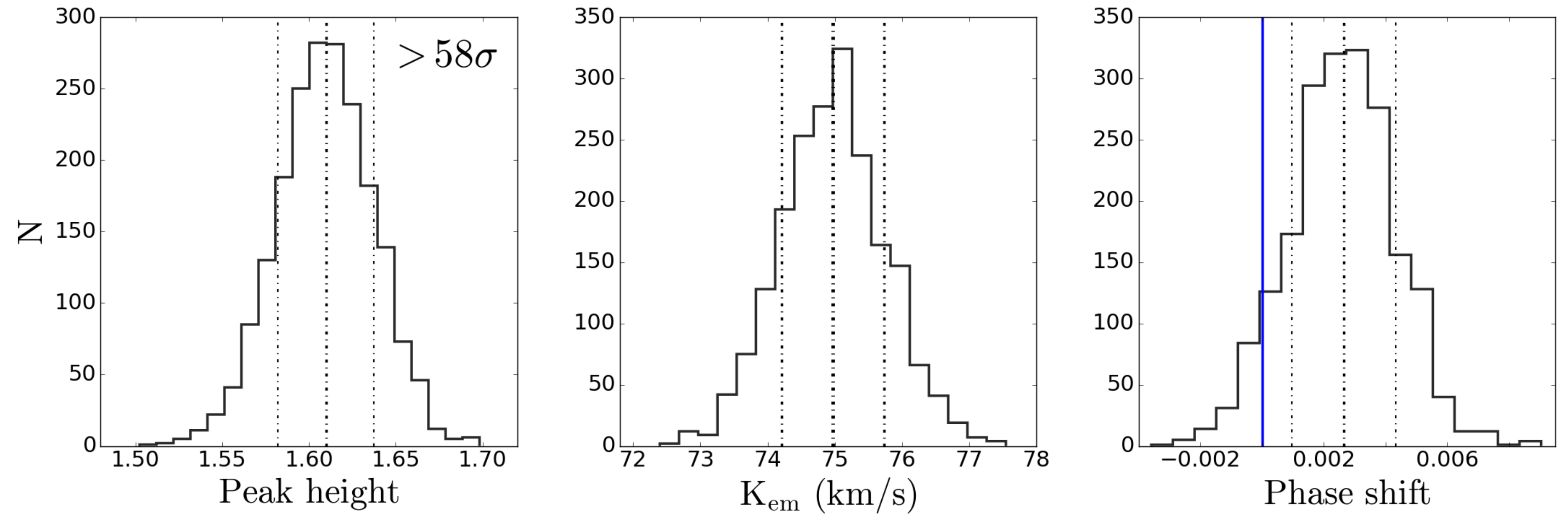}
      \caption{Number distributions of the peak height (left), RV semi-amplitude (middle) and phase shift (right) of the donor emission spot determined from 2000 combined, bootstrapped images. Dashed lines indicate the mean and the $\pm$ 1$\sigma$ confidence intervals. The feature is significant at the >58$\sigma$ level and the phase shift is consistent with zero (blue solid line) within 2$\sigma$ using the orbital period, ephemeris and systemic velocity in Table~\ref{tab:scoparams}.}
    \label{fig:boothist}
\end{figure*}

Fig.~\ref{fig:boothist} shows the histograms of the most relevant spot parameters as derived from the bootstrap samples.
Firstly, we considered the distribution of the peak height in terms of a significance level of the (combined) donor signal.
By fitting a Gaussian curve to the histogram plot in the left panel of Fig.~\ref{fig:boothist}, we deduce that the mean of the peak intensity distribution is above zero at the >58$\sigma$ level.
Moreover, in contrast to the results for the much lower SNR data of XTE J1814$-$338 (which indicated a 4$\sigma$ detection of the mass donor), Fig.~\ref{fig:boothist} (middle panel) shows that the combined Bowen image for Sco X-1 provides a tight constraint on the RV semi-amplitude, $\mathrm{K_{em}}$ = 75.0 $\pm$ 0.8 km $\mathrm{s^{-1}}$, for an assumed value of $\gamma$ = -113.6 km $\mathrm{s^{-1}}$.
 
It is known that the choice of the input parameter $\gamma$ can have an impact on the resulting reconstruction -- and spot features in particular would appear out-of-focus and misplaced if an incorrect $\gamma$ was used to obtain the image (e.g. \citealt{2003MNRAS.344..448S}; \citealt{2009MNRAS.394L.136M}).
Thus, we ran the same analysis using a range of $\gamma$ values in steps of 2 km $\mathrm{s^{-1}}$, centered on $-$113.6 km $\mathrm{s^{-1}}$.
Within a narrow range of $\gamma$ values between $-$120 and $-$108 km $\mathrm{s^{-1}}$, a >50$\sigma$ donor detection was achieved, peaking near the expected value from our RV fit. The derived $\mathrm{K_{em}}$ velocity showed a small range of variation (with a maximum of 0.6 km $\mathrm{s^{-1}}$ deviation from 75 km $\mathrm{s^{-1}}$), therefore indicating the presence of a small systematic error on $\mathrm{K_{em}}$ due to the uncertainty in $\gamma$. We note that this constraint on the systemic velocity ($\gamma=-114\pm6$ km $\mathrm{s^{-1}}$) is looser compared to the best-fit value derived from the RVs ($-113.6\pm0.2$ km $\mathrm{s^{-1}}$) since it is not very sensitive to changes below the spectral resolution of the data.
Based on the above results, we obtain
\begin{center}
$\mathrm{K_{em,\, doppler}}$ = 75.0 $\pm$ 0.8 (stat) $\pm$ 0.6 (sys) km $\mathrm{s^{-1}}$, \\
\end{center}
in excellent agreement with the value derived from the traditional RV fitting method.
Therefore the novel technique of bootstrap Monte Carlo delivers a robust and more conservative estimate of $\mathrm{K_{em}}$ for the combined observations, which we adopt as our best estimate of the Bowen $\mathrm{K_{em}}$ amplitude.
Since Doppler tomography exploits all spectra at once, the technique is also less affected by the problem of extremely weak features near phase zero (see G14; Fig. 2) compared to individual line profile fitting.
This makes bootstrap Doppler tomography an attractive tool for deriving binary system parameters \emph{across all SNRs}.
The main disadvantage of the tomography-based method is that $T_0$ and $\mathrm{P_{orb}}$ are used as input, instead of being determined directly from the reconstruction.
However, the phase shift ($\Delta$$\mathrm{\phi_{spot}}$) between the donor spot and the vertical plane can provide a good indication of the accuracy of the input ephemeris.
 
With the correct ephemeris, emission from the donor should always be projected on the tomogram at the position $V_x$ = 0 and $V_y$ = $\mathrm{K_{em}}$ (hence $\Delta$$\mathrm{\phi_{spot}}$ = 0), whereas phase errors tend to lead to a rotation of the image.
In the right panel of Fig.~\ref{fig:boothist} we show that $\Delta$$\mathrm{\phi_{spot}}$ is consistent with zero (blue solid line) within 2$\sigma$ 
for our best estimate of $\gamma$.
A diagnostic plot of $\Delta$$\mathrm{\phi_{spot}}$ as a function of the assumed $\gamma$ (between -120 and -108 km $\mathrm{s^{-1}}$) gives
$\Delta$$\mathrm{\phi_{spot}}$ = 0.003 $\pm$ 0.002 (statistical) $\pm$ 0.002 (systematic).
It can thus be concluded that the donor spot is indeed centered on the positive $V_y$-axis
using the revised binary period and ephemeris.
Table~\ref{tab:scoparams} lists the final adopted values of $\mathrm{T_0}$, $\mathrm{P_{orb}}$, $\gamma$, $\mathrm{K_{em}}$ and the cross-term (needed to determine the propagated uncertainty on the future epoch of inferior conjunction; see G14).

A similar test using $\mathrm{T_0}$ and $\mathrm{P_{orb}}$ of G14 yielded a significant (3$\sigma$) phase shift between the donor spot and the vertical plane, which indicates that the previous results included contributions from systematic errors.
Since the revised period and ephemeris were derived from an improved RV curve (including all data points) using a refined analysis approach, and there is no measurable phase shift,  
the new estimates for orbital parameters presented in this paper can be considered more reliable than the results of G14.
 
\section{Estimation of binary parameters}
\label{sec:discuss}
So far we have exploited the proven Bowen emission line diagnostic that permits a robust and accurate RV study of the companion star in Sco X-1. 
The main caveat is that the narrow lines are produced on the irradiated inner hemisphere of the secondary, instead of the entire Roche lobe; hence the previously determined Bowen $\mathrm{K_{em}}$ amplitude represents only a lower limit to the true, centre-of-mass velocity of the companion, $K_2$.
Therefore a `K-correction' must be applied to the observed $\mathrm{K_{em}}$ when performing binary parameter calculations.
 
To precisely constrain the remaining parameters for the purpose of GW searches, $K_1$ and $e$, requires a direct determination of the neutron star orbit, which is particularly challenging due to the lack of NS pulsations.
While coherent pulsations of Sco X-1 are not observed,
we shall rely on the known set of parameter constraints and apply Kepler's laws to derive a more conservative $K_1$ estimate compared to that of SC02 obtained using broad disc emission features.
 
\subsection{The K-correction}
\label{sec:kcorrection}
The deviation between the reprocessed light centre and the centre of mass of the Roche-lobe-filling donor ($\mathrm{K_{cor}}$ = $\frac{\mathrm{K_{em}}}{K_2}$ $<$ 1) can be modelled by generating RV curves from simulated emission line profiles formed by isotropic irradiation on the illuminated hemisphere.
Results of the most extensive study of K-correction modelling to date (\citealt{2005ApJ...635..502M}; hereafter, MCM05) suggest a weak dependence of $\mathrm{K_{cor}}$ on the orbital inclination ($i$); but a strong dependence on the mass ratio between the donor star and the compact object ($q$ = $M_2$/$M_1$) and the disc opening angle ($\alpha$), which represents the shielding effect by a flared and opaque accretion disc.
By fitting fourth-order polynomial models to the synthetic K-corrections (as functions of $q$), MCM05 derived parametric approximations to $\mathrm{K_{cor}}$ = $\frac{\mathrm{K_{em}}}{K_2} \cong N_0 + N_1 q + N_2 q^2 + N_3 q^3 + N_4 q^4$, for different disc flaring angles between $0^{\circ}$ and $18^{\circ}$, in both low ($i$ = $40^{\circ}$) and high inclination ($i$ = $90^{\circ}$) cases (see Fig. 4 and Table 1 of their paper).
 
For Sco X-1, an estimate of the binary inclination is available from modelling of the orientation of radio lobes ($i$ = $44^{\circ}$ $\pm$ $6^{\circ}$; \citealt{2001ApJ...558..283F}).
However, the disc shielding parameter is unknown and therefore throughout the rest of our analysis, we consider the full range of possibilities -- from the extreme case of neglecting disc shadowing ($\alpha$ = $0^{\circ}$; maximum displacement) to $\alpha$ = $18^{\circ}$ where the companion is nearly completely shadowed by the accretion disc (minimum displacement).
The key ingredient for improving the K-correction then involves mainly the determination of constraints on $q$.
 
\subsubsection{Center of symmetry search}
\label{sec:COS}
The ring-like disc emission feature revealed using Doppler tomography (SC02 and this paper) may be used to constrain the NS's projected RV ($K_1$).
In Doppler coordinates, the projection of an ideal Keplerian disc is always centered on the expected position of the NS ($V_x$, $V_y$) = (0, -$K_1$), as the disc gas should track the orbital motion of the accreting primary.
Thus, although the accretor is rarely seen in emission, a search for the centre of symmetry (CoS) of the disc component in velocity space should, in principle, yield a valid solution for $K_1$ $\simeq$ $\mathrm{K_{disc}}$ (e.g. \citealt{2010MNRAS.401.1857V}; \citealt{2015MNRAS.447..149L}).

For a reconstructed Doppler map, one first subtracts a symmetric component around an assumed centre ($V_x$ = 0, $V_y$ = -$\mathrm{K_{disc}}$),
followed by inspecting the resulting residual image for any remaining asymmetries not related to the primary, e.g., emission caused by the stream/disc impact. 
The process is then repeated for a wide range of $\mathrm{K_{disc}}$ values to select an optimal center that minimizes the residuals over a region of the map containing clean disc emission.
Since non-disc components and hot spot contamination can be easily masked during the minimization, the method has the advantage of allowing one to remove the distortions by any asymmetries which would normally disrupt the outer disc, and focus on the higher velocity emission arising from the (undisturbed) inner part of the disc (appearing further away from the origin in velocity space).
 
Under the assumption that the disc is axisymmetric around the NS primary, the technique was adopted in SC02 to obtain an initial $K_1$ estimate for Sco X-1 based on the location of the CoS for the H$\beta$ image ($\mathrm{K_{disc,\, H\beta}}$ = 40 km $\mathrm{s^{-1}}$).
Here we focus on applying the method to the newly computed Bowen image -- which reveals the best defined accretion disc among the maps reconstructed from principal spectral features (see, e.g., SC02; Fig. 4) -- in order to assess the stability and the validity of the CoS measurement.
 
We cycled the assumed $V_y$-coordinate of the disc centre between -200 and +20 km $\mathrm{s^{-1}}$ in steps of 2 km $\mathrm{s^{-1}}$ (while keeping $V_x$ fixed to 0) and measured the mean residuals in the lower half of the reconstruction (to exclude the region near the secondary emission).
For the 1999 Bowen data the residuals showed a distinct minimum at $V_y$ = -$\mathrm{K_{disc}}$ = -84 km $\mathrm{s^{-1}}$.
A similar result was obtained also for the 2011 Bowen image ($\mathrm{K_{disc}}$ = 90 km $\mathrm{s^{-1}}$), suggesting that a significantly higher value of $K_1$ amplitude than previously estimated (using the same method but a different line feature) needs to be considered.
Given the absence of a sharp disc ring in the 1999 H$\beta$ map, compounded by the presence of enhanced emission in the bottom left quadrant,
it is most likely that the best fit $\mathrm{K_{disc,\, H\beta}}$ value still failed to represent the NS's true projected RV.
These significant differences between different lines show that extracting robust $K_1$ constraints from disk lines remains challenging. 
Hence we decided to 
adopt the greatest value for $\mathrm{K_{disc}}$ based on CoS measurements using maps of a range of emission lines as our most conservative maximum value of $K_1$.
Using a new upper limit on
\begin{center}
$K_1$ $\simeq$ $\mathrm{K_{disc,\, Bowen}}$ $\leqslant$ 90 km $\mathrm{s^{-1}}$, 
\end{center}
and apply the K-correction (for $\alpha$ = $18^{\circ}$) to $\mathrm{K_{em}}$, we obtain an upper limit of $q$ $\leq$ 0.76.
 
\subsubsection{Rotational line broadening}
Assuming rotation to be the dominant line broadening mechanism, the observed width of the sharp \ion{N}{iii}/\ion{C}{iii} lines ($\mathrm{V_{rot}}$ sin$\emph{i}_{\mathrm{em}}$) to $K_2$ can be used to set a strict lower limit on $q$. For a spectral feature that is produced throughout the Roche lobe, the expected width ($\mathrm{V_{rot}}$ sin$\emph{i}$) is given by the relation,
\begin{equation}
\label{eq:Vsini}
\frac{\mathrm{V_{rot}}\,    \mathrm{sin}\emph{i}}{K_2} \simeq 0.462\,   \big[(1+\emph{q})^2 q\big]^{\frac{1}{3}}
\end{equation}
\citep{1988ApJ...324..411W}. Since the donor is not fully illuminated, this becomes a limit as $\mathrm{V_{rot}}$ sin$\emph{i}$ > $\mathrm{V_{rot}}$ sin$\emph{i}_{\mathrm{em}}$.
By taking the weighted average of the set of 12 FWHM (\ion{N}{iii}/\ion{C}{iii}) values measured from the VLT spectra (see Section~\ref{sec:sourcedata}) near orbital phase 0.5 when the visibility of the irradiated face is maximum,
we obtain an estimate for the width of the narrow components of FWHM = 51.9 $\pm$ 0.5 km $\mathrm{s^{-1}}$.
However, it should be cautioned that this preliminary estimate does not provide a strict lower limit to $\mathrm{V_{rot}}$ sin$\emph{i}$ as the measured widths include also the intrinsic broadening effect due to our instrumental resolution (8 km s$^{-1}$).
 
Following \citet{2006MNRAS.373.1235C}, we corrected for the instrumental effect by broadening a Gaussian template of FWHM = 8 km s$^{-1}$ using a Gray rotational profile \citep{1992oasp.book.....G} without limb-darkening (because fluorescence lines arise in optically thin conditions); and
subsequently determined for each of the 12 VLT measurements the amount of rotational broadening required to reproduce the apparent FWHM (\ion{N}{iii}/\ion{C}{iii}).
This leads to a weighted average of $\mathrm{V_{rot}}$ sin$\emph{i}_{\mathrm{em}}$ = 37.8 $\pm$ 0.4 km $\mathrm{s^{-1}}$, which 
combined with the K-correction for $\alpha$ = $0^{\circ}$, gives a secure lower limit of $q$ $\geq$ 0.22.

The lower limit of $q$ can be further improved by applying a correction factor (defined as $V$$\mathrm{sin\emph{i}_{cor}}$ = $\frac{\emph{V}\mathrm{sin}\emph{i}_\mathrm{{em}}}{\emph{V}\mathrm{sin}\emph{i}}$ $<$ 1) to convert the observed rotational broadening to true $V$sin$\emph{i}$, for the reasons mentioned above.
In light of the results of the MCM05 study of $\mathrm{K_{cor}}$, 
we estimate this correction factor $V$$\mathrm{sin\emph{i}_{cor}}$ in the low inclination case ($i$ = $40^{\circ}$) -- suitable to be applied to the case of Sco X-1 -- by following the same approach for modelling the K-correction described in MCM05.
The details of the modelling are described in the appendix.  
The solution for $V$$\mathrm{sin\emph{i}_{cor}}$ at $\alpha$ = $0^{\circ}$ represents the most conservative correction (see appendix), which can be safely adopted to yield an improved lower limit of $q$ (at a value slightly higher than 0.22, Section \ref{sec:sco-monte}), and thereby also providing the best lower limit to $K_1$ (see Section \ref{sec:sco-k1}). 


\subsection{Monte-Carlo analysis}
\label{sec:sco-monte}
The discovery of the Bowen emission lines (made more than a decade ago) arising from the donor star in Sco X-1, combined with the inclination derived by \citet{2001ApJ...558..283F}, led to some of the first dynamical parameter constraints for this prototypical LMXB (SC02).
In this section, armed with the complete set of K-correction models solved by MCM05 and new limits on $K_1$ and the rotational broadening (see Section~\ref{sec:kcorrection}), we perform Monte Carlo simulations for deriving the probability density functions (PDF's) of Sco X-1 system parameters. Of particular note is the $K_1$ distribution (or $a_x$sin$i$ = $K_1$$\mathrm{P_{orb}}$/(2$\pi$) ) required by most directed searches for continuous GWs.

As an initial, conservative estimate, we used the binary inclination inferred from the orientation of radio jets and results from the reanalysis of the combined Bowen data (Section~\ref{sec:analysis} only). Synthetic values of $\mathrm{P_{orb}}$, $\mathrm{K_{em}}$ and $i$ were selected from a Gauss-normal distribution (with the mean and standard deviation equal to the values listed in Table~\ref{tab:scoparams}).
To perform the K-correction, we started with the most conservative scenario by simulating uniformly-distributed values of $\alpha$ and $q$ over the entire ranges considered by MCM05 (covering values typical of XRBs) between $\alpha$ $\sim$ $0^{\circ}$ and $18^{\circ}$; $q$ $\sim$ 0.05 and 0.83.
For each pair of randomly generated disc opening angle and mass ratio, we determined a precise value of $\mathrm{K_{cor}}$ by interpolating over the grid of MCM05 models.
After applying $\mathrm{K_{cor}}$ to $\mathrm{K_{em}}$, we obtained the corresponding $K_2$ amplitude, which was then
used (along with $\mathrm{P_{orb}}$, $i$ and $q$) as the input to the mass function equation
\begin{equation}
f(M) = \frac{K_2^3\rm~P_{orb}}{2\pi G} = \frac{M_1\rm~sin^3(\emph{i})}{(1+q)^2}
\label{eq:massfunc} 
\end{equation} 
for deriving the mass of the compact primary $M_1$.
At the end of each trial, the RV of the primary $K_1$ = $q$$K_2$ and the mass of the donor $M_2$ were also calculated; and the process was repeated $10^6$ times.
Assuming a NS accretor, only the outcomes of trials that yielded an $M_1$ value between the minimum ($\sim$ 0.9 $M_{\odot}$; e.g., \citealt{2012ARNPS..62..485L}) and the maximum stable NS mass (3.2 $M_{\odot}$) were assembled into the initial, conservative estimate of the PDF's (Figs~\ref{fig:K_contour} and~\ref{fig:M_contour}; grey).

\subsubsection{Radial velocity of the NS}
\label{sec:sco-k1}
\begin{figure}
        \includegraphics[width=\columnwidth]{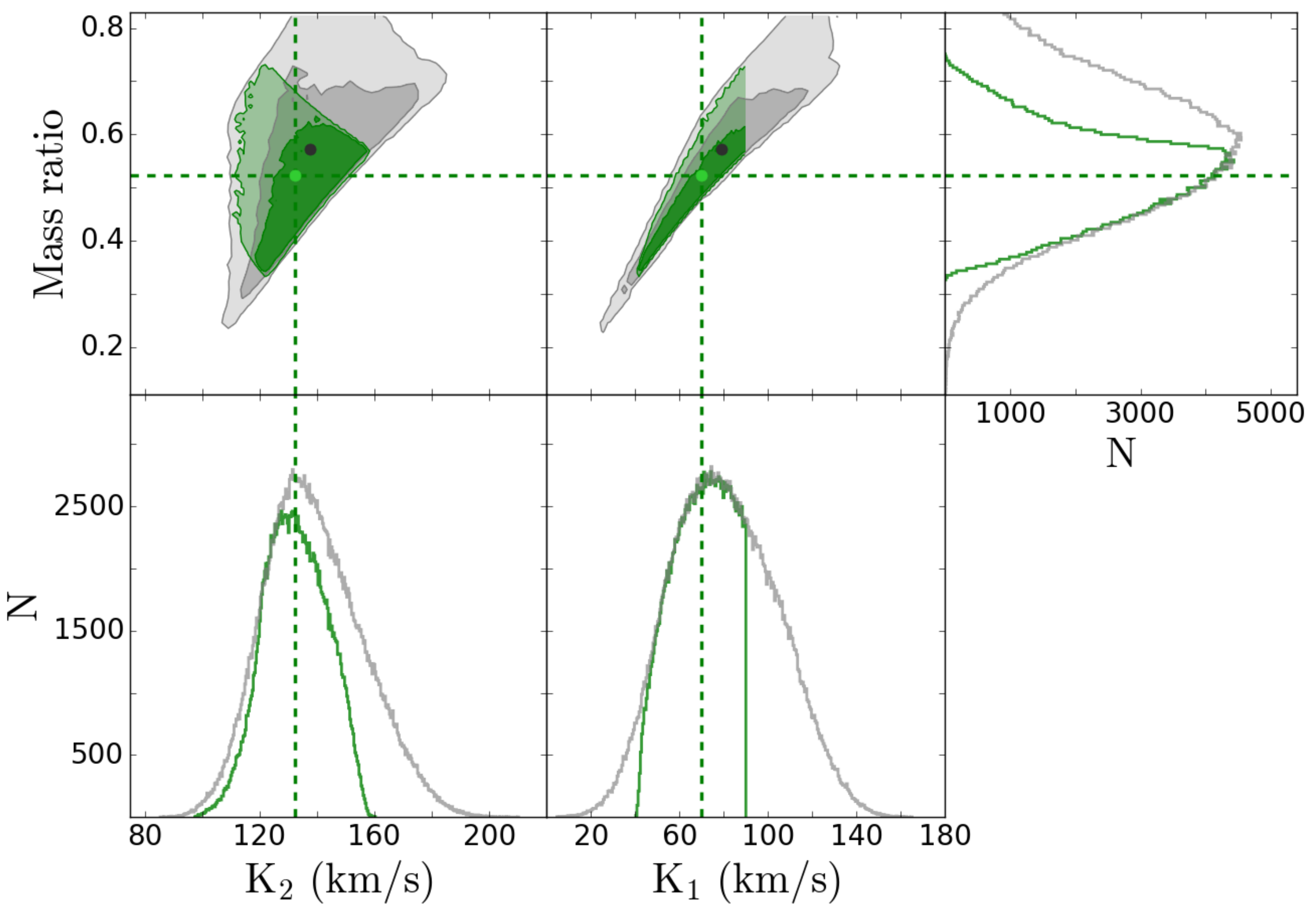}
    \caption{The 68 per cent and 95 per cent confidence regions and projected PDF's for $K_1$, $K_2$ and $q$ derived from the Monte Carlo binary parameter calculations ($10^6$ trials). Our best estimates for the PDF's (after applying the newly derived $K_1$ and $V$$\mathrm{sin\emph{i}}$ constraints; see Section~\ref{sec:kcorrection}) are shown in green, compared to the results of the most conservative possible scenario (grey). Dashed lines indicate the locations of the $\mathrm{50^{th}}$ percentile (median). The best estimates for the system parameters and their associated 95 per cent confidence intervals (CI) error bars are given in Table~\ref{tab:scoparams}.}
    \label{fig:K_contour}
\end{figure}

\begin{figure}
        \includegraphics[width=\columnwidth]{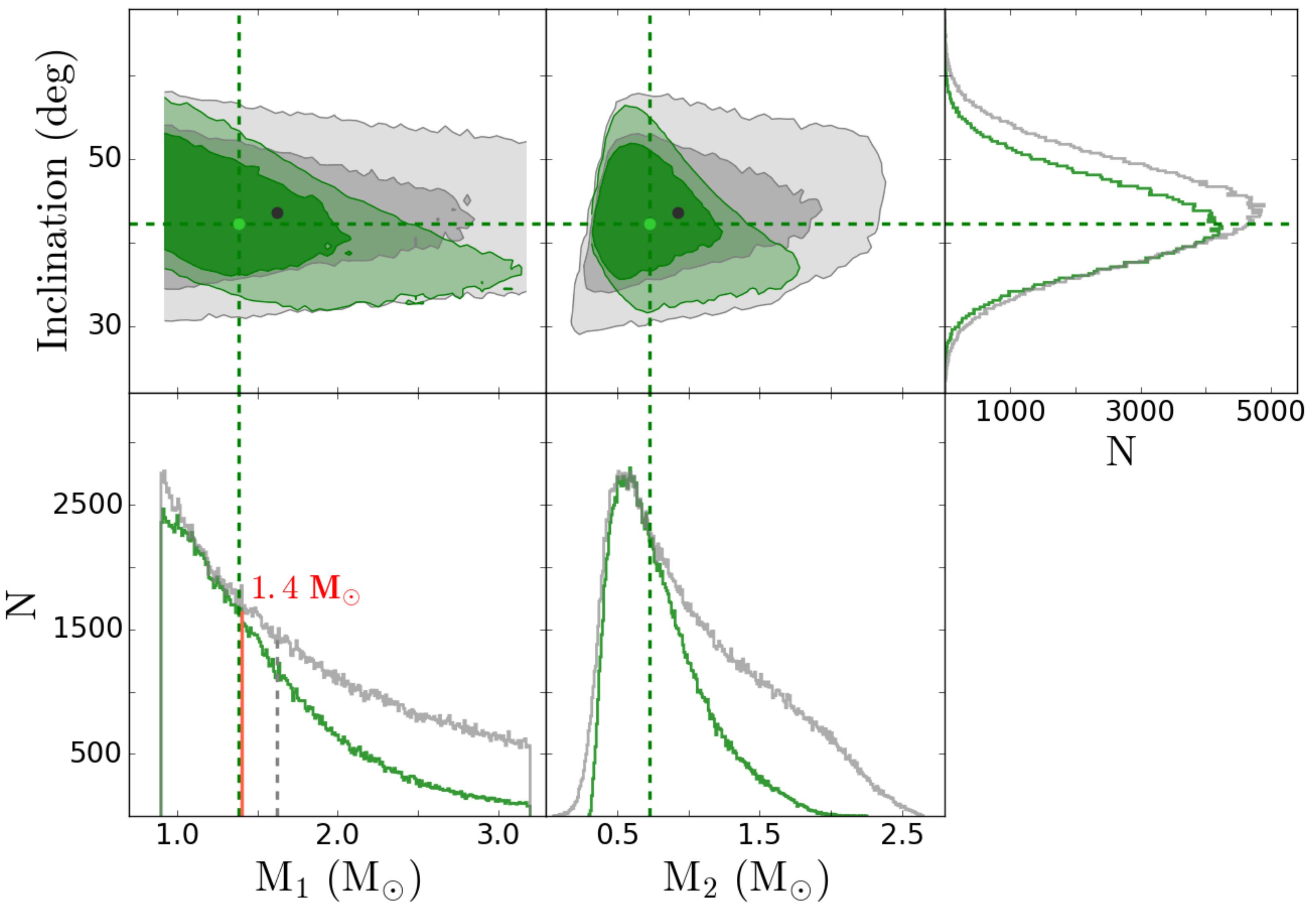}
    \caption{The 68 per cent and 95 per cent confidence regions and projected PDF's for $M_1$ and $M_2$ using the inclination estimate by \citet{2001ApJ...558..283F}. Dashed lines indicate the locations of the $\mathrm{50^{th}}$ percentile (median). Note that including the observational constraints on $K_1$ and $V$$\mathrm{sin\emph{i}}$ in the Monte Carlo analysis shifts the median exactly to the canonical NS mass of 1.4 $M_{\odot}$ (red solid line).}
    \label{fig:M_contour}
\end{figure}

As a next step we imposed an upper limit on $K_1$ determined from center of symmetry analyses of the disc component
($K_1$ $\lesssim$ 90 km $\mathrm{s^{-1}}$), which essentially sets a more realistic upper limit of $q$ as discussed in Section~\ref{sec:COS}.
For each Monte Carlo trial, we additionally computed the minimum allowable $V$$\mathrm{sin\emph{i}}$ value by applying the most conservative $V$$\mathrm{sin\emph{i}}$-correction (derived in the appendix). This is the case of $\alpha$ = $0^{\circ}$, dependent only on $q$, and considering the lower limit on $V$$\mathrm{sin\emph{i}}$ $>$ 36.6 km s$^{-1}$ (99.7 per cent confidence) as estimated from the width of the sharp \ion{N}{iii}/\ion{C}{iii} lines.
A conservative and stringent lower limit of $q$ could therefore be established using the relationship between $V$$\mathrm{sin\emph{i}}$ and $q$ (equation~\ref{eq:Vsini}).
Finally, the output PDF's that reflect our current best knowledge of Sco X-1 parameters are shown in Figs~\ref{fig:K_contour} and~\ref{fig:M_contour} (green), with those derived only from the observed distributions of $\mathrm{P_{orb}}$ and $\mathrm{K_{em}}$.

Because of these newly imposed constraints,
the CIs for nearly all system parameters (except $M_1$) are narrowed down significantly. 
In particular, the estimate of the centre of the disc component had such a strong impact
that the upper $K_1$ limit was reduced from 165 to 90 km $\mathrm{s^{-1}}$;
and thus represents the most valuable determination in the context of refining the parameter space for GW searches. Ruling out high values of $K_1$ (or $a_x$sin$i$) has a double benefit, since high values of $a_x$sin$i$ increase the required resolution in orbital phase ($T_0$).
We also note that the new $\mathrm{V_{rot}}$ sin$\emph{i}$ constraint (combined with a minimum $V$$\mathrm{sin\emph{i}}$-correction) provides a refined lower limit not only for $q$ but also for both $K_1$ and $K_2$, due to the dependence of $\mathrm{K_{cor}}$ on $q$. 
This yielded the current best hard limits on $K_1$ of 40--90 km $\mathrm{s^{-1}}$, or equivalently, 
\begin{center}
1.45 ls $\leq$ $a_x$sin$i$ $\leq$ 3.25 ls.
\end{center}
This range of $K_1$ is significantly broader than the previous constraint (40 $\pm$ 5 km $\mathrm{s^{-1}}$) derived solely from centre of symmetry measurements from the H$\beta$ Doppler map (SC02). 
In the recent search for continuous gravitational-waves from Sco X-1 (e.g., \citealt{2017ApJ...847...47A}), the range for the projected semi-major axis has been expanded in order to cover the full parameter space.

\subsubsection{Determination of the component masses}
\label{sec:masses}
Despite the fact that the RV semi-amplitude of the Bowen emission can be measured with a precision of $\sim$1 per cent, 
the true projected RV of the donor is not well constrained ($K_2$ = $132^{+20}_{-21}$ km $\mathrm{s^{-1}}$; 95 per cent confidence) given the lack of knowledge of $q$ and the disc shielding parameter.
Using a geometric model for the reprocessing of X-rays in LMXBs, an average disc opening angle of $\sim$$12^{\circ}$ has been proposed for other persistent sources, e.g., GR Mus and V4134 Sgr \citep{1996A&A...314..484D}.
Similarly, for Sco X-1, an $\alpha$-value at the lowest end of the range between $0^{\circ}$ and $18^{\circ}$ is perhaps unlikely (given also the clear presence of an accretion disc). 
However, we chose 
to consider all $\alpha$ values within the most conservative range between 0--$18^{\circ}$ in the Monte Carlo analyses to avoid introducing biases into the resulting PDF's.

Due to the large uncertainties in the final values of $K_2$ (dominated by the unknown K-correction) and $q$ ($0.52^{+0.16}_{-0.15}$; 95 per cent),
it remains difficult to put strong constraints on the component masses.
Using the inclination value of $44^{\circ}$ $\pm$ $6^{\circ}$ -- inferred from the orientation of radio jets 
(assuming that the radio jet is perpendicular to the orbital plane) -- we obtain a 95 per cent credible range of the NS mass between 0.9--2.8 $M_{\odot}$ and an $M_2$ range of 0.4--1.5 $M_{\odot}$ (95 per cent).
Interestingly though, the median of our best estimate for the $M_1$ PDF (after applying the $K_1$ and $V$$\mathrm{sin\emph{i}}$ constraints) coincides with the `canonical' value of 1.4 $M_{\odot}$ -- in support of the presence of a $\sim$ 1.4 $M_{\odot}$ neutron star.

The complete set of dynamical system parameter constraints, derived from the latest RV measurements as well as Monte-Carlo simulations, are provided in Table~\ref{tab:scoparams}.

\begin{table}
        \centering
        \caption{Sco X-1 system parameters. Note that the $a_x$sin$i$ constraints represent hard limits on the projected semi-major axis of the NS.}
        \label{tab:scoparams}
        \begin{tabular}{l l}
                \hline
                Parameter &  Value \\
                \hline
                $\mathrm{P_{orb}}$ $\mathrm{(d)}$ & 0.7873132 $\pm$ 0.0000005  \\
                $T_0$$\rm~HJD(UTC)$ & 2455522.6668 $\pm$ 0.0006 \\
                $T_0$ $\mathrm{(GPS}$ $\mathrm{seconds)}$ & 974433630 $\pm$ 50 \\
                $|$$V$($\mathrm{P_{orb}}$, $T_0$)$|$ $\mathrm{(d^2)}$ & 3 $\times$ $\mathrm{10^{-14}}$ \\
                $\gamma$ (km s$^{-1}$) & -113.6 $\pm$ 0.2 \\
                $\mathrm{K_{em}}$ (km s$^{-1}$) & 75.0 $\pm$ 0.8 (statistical) $\pm$ 0.6 (systematic) \\
                \noalign{\vskip 1mm}
                $K_{1}$$(\rm~$km$\rm~s^{-1})$ & $70^{+19}_{-25}$ (95\%) \\
                \noalign{\vskip 1mm} 
                $a_x$sin$i$ (ls) & $[1.45, 3.25]^a$ \\
                \noalign{\vskip 1mm}
                $K_{2}$ (km s$^{-1}$) & $132^{+20}_{-21}$ (95\%) \\
                \noalign{\vskip 1mm}
                $q$ ($M_2$/$M_1$) & $0.52^{+0.16}_{-0.15}$ (95\%) \\
                \noalign{\vskip .5mm}
                $i$ $(^{\circ})^b$ & 44 $\pm$ 6 \\
                \noalign{\vskip .5mm}
                $M_1$$\rm~(M_{\odot}$) & $1.4^{+1.4}_{-0.5}$ (95\%) \\
                \noalign{\vskip 1mm}
                $M_2$$\rm~(M_{\odot}$) & $0.7^{+0.8}_{-0.3}$ (95\%) \\
                \hline
        \end{tabular}
\newline
\raggedright \emph{Notes.} $\mathit{^a}$The range for the projected semi-major axis of the NS $a_x$sin$i$ = $K_1$$\mathrm{P_{orb}}$/(2$\pi$) in light-seconds; $\mathit{^b}$Adapted from \citep{2001ApJ...558..283F}.
\end{table}

\subsection{The eccentricity problem}
\label{sec:aE} 
\begin{figure}
        \includegraphics[width=\columnwidth]{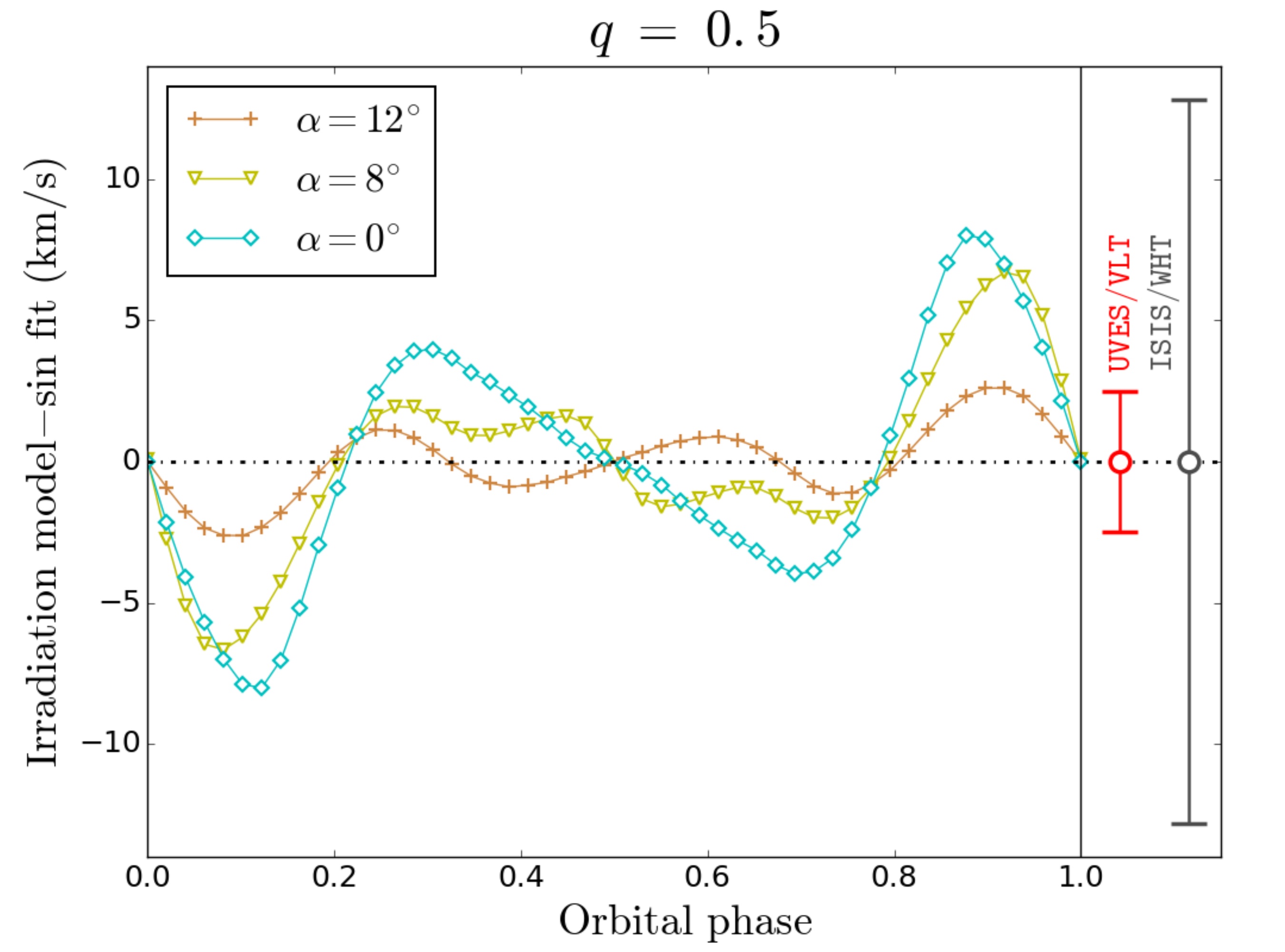}
    \caption{Plot of simulated residual RVs obtained by subtracting the sinusoidal fit from the irradiation model (computed in the appendix) for a probable mass ratio of Sco X-1 ($q$ = 0.5) and three different disk flaring angles. The grey and red error bars represent respectively the average (rescaled) RV uncertainties for the WHT and UVES/VLT measurements (derived in Section~\ref{sec:sinfit}). Neither WHT nor the highest resolution UVES/VLT data could offer a high enough RV precision to firmly differentiate between residual RVs for low and high $\alpha$ values,
which makes it infeasible to clearly disentangle any spurious eccentricity produced by irradiation effects from the true orbital eccentricity (in the case of Sco X-1).} 
    \label{fig:lprofile_resid}
\end{figure}

One complication of the Bowen technique is that the established RV curve is not expected to be perfectly sinusoidal owing to the effect of non-uniform flux distribution at the surface of the irradiated donor star. This is a particular handicap for attempts to constrain any true orbital eccentricity,$e$, of the binary.
The influence of irradiation effects can be clearly seen in Fig.~\ref{fig:lprofile_resid}, where we display the simulated residual RV curves obtained by subtracting the sinusoidal fit from the irradiation model (described in the appendix) for the most probable mass ratio of Sco X-1 ($q$ $\simeq$ 0.5; as calculated by the Monte Carlo method) and different $\alpha$-values.
As noted in the appendix and in previous studies, the amount of deviation of the simulated RVs from a sinusoidal wave form increases with decreasing disc opening angle (see also MCM05; Fig. 2).
Closer inspections reveal that for all $\alpha$ values the deviations increase at small and large orbital phases
($\lesssim$ 0.2 and $\gtrsim$ 0.8),
consistent with the distortions as seen, for example, in the observed RV curves of the pre-cataclysmic binary NN Ser (see \citealt{2010MNRAS.402.2591P}; Fig. 8).
Based on results from both observations and simulations, we conclude that RV distortions caused by irradiation effects would 
dominate those produced by any small intrinsic $e$ of the binary orbit. Although they are not described by an eccentric model, attempting to fit such RV curves with an eccentric orbit would inevitably produce an \emph{apparent eccentricity} ($a$E), which varies strongly with disc opening angle\footnote{As we mentioned in the appendix, the amount of deviation of RVs from a pure sinusoid increases with increasing mass ratio, therefore $a$E varies also with $q$.}.

Worse still, Fig.~\ref{fig:lprofile_resid} also shows that 
the WHT data do not yet offer the RV precision high enough for differentiating between the irradiation and the circular orbit model.
Not even the 2011 UVES/VLT data has the precision high enough to firmly differentiate between residual RVs for low and high $\alpha$ angles, 
thus making it currently infeasible to accurately infer the value of the disc opening angle necessary to clearly disentangle $a$E from any true orbital eccentricity. 

Due to the inherent limitations of the method, as well as limitations of the current data sets,
in this paper we only attempt to place a crude upper bound on $e$
by running an MCMC analysis (to fit the combined RV data) 
and assuming an elliptical orbit model.
By leaving the spurious eccentricity ($E$) and periastron angle as free parameters,
we obtained a 95 per cent credible range of the eccentricity parameter $E$ between 0.00086-0.082 (using a uniform prior in log $E$).
However, we stress that the nominal upper bound of 0.082, determined
\emph{without} taking account of any irradiation effects (note that $a$E/$e$ $\gg$ 1), 
cannot accurately represent the limit to the true orbital $e$, which is expected to be much closer to zero for a Roche-lobe overflow XRB with a high mass-accretion rate (e.g., Sco X-1).



\section{Conclusions and Future Perspectives}
We presented a reanalysis of the combined spectroscopic data of SC02 and G14 with the aim of providing revised constraints on key orbital parameters of Sco X-1. Here we paid particular attention to the requirements of directed searches for continuous GWs.
By exploiting the previously established spectroscopic diagnostic of the mass donor, we derived an improved RV curve of the Bowen blend, arising from the X-ray heated hemisphere of the companion, and demonstrated that both the binary period and ephemeris can be precisely determined using the traditional RV fitting method.
Additionally, Doppler tomography of the Bowen emission region clearly reveals a donor spot feature along the positive $V_y$-axis, providing a viable alternative to reliably measure the RV semi-amplitude $\mathrm{K_{em}}$ (< $K_2$).
In the absence of coherent X-ray pulsations,
a precise measurement of the projected RV semi-amplitude of the NS $K_1$ is still challenging.

The latest centre of symmetry measurements (from the Bowen Doppler tomogram) yielded a CoS of a sharp disc component of 90 km $\mathrm{s^{-1}}$,
which we interpret as a revised upper limit to $K_1$ (larger than previously reported by SC02).
We explore the implications of these by implementing Monte Carlo binary parameter calculations, combined with the K-correction (to transform $\mathrm{K_{em}}$ to $K_2$)
and the rotational broadening $V$$\mathrm{sin\emph{i}}$.
By considering a random distribution of mass ratios and disc opening angles, we obtain a looser constraint of $K_1$ (40--90 km $\mathrm{s^{-1}}$, or equivalently, 1.45 ls $\leq$ $a_x$sin$i$ $\leq$ 3.25) than that derived in SC02 (40 $\pm$ 5 km $\mathrm{s^{-1}}$).
Finally, we discussed the pitfalls of measuring the eccentricity of the NS orbit;
namely, irradiation effects are known to produce an apparent eccentricity, which is hard to disentangle from the true orbital $e$ using the current Bowen dataset.
By simply allowing for an ellipticity in the fit to the RV curve (without taking account of any irradiation effects),
a nominal upper limit for the eccentricity $E$ < 0.082 was obtained.
However, we stress that the true $e$ is expected to be much closer to zero in the case of Sco X-1, and has therefore been ignored in recent directed searches with (advanced) LIGO data (e.g. \citealt{PhysRevD.95.122003}; \citealt{2017ApJ...847...47A}).  


In light of our new constraints on orbital parameters, 
the ranges of search parameters $T_0$ and $\mathrm{P_{orb}}$ should be updated with the refined period and ephemeris. 
More importantly, the range for the projected semi-major axis $\mathrm{a_x}$sin$i$ needs to be expanded and updated with the new confidence interval for $K_1$, which reflects the systematic uncertainty much greater in magnitude than the statistical error estimated in SC02.
With the previous values (which spanned an unrealistically narrow range), there was a high probability of missing the GW signal.
In the first Advanced LIGO observing run (O1) searches, a preliminary constraint of 10 km $\mathrm{s^{-1}}$ $\leq$ $K_1$ $\leq$ 90 km $\mathrm{s^{-1}}$ (available at the time the search was constructed) was adopted as the prior assumption. 
The best marginalized 95\% upper limit on the signal amplitude (obtained by \citealt{2017ApJ...847...47A}) reaches  
2.3 $\times$ $\mathrm{10^{-25}}$,
which is a factor of $\sim$ 7 stronger than the best upper limits set using data from initial LIGO science runs.
In future runs, computing resources will be concentrated on the final refined range of 40 km $\mathrm{s^{-1}}$ $\leq$ $K_1$ $\leq$ 90 km $\mathrm{s^{-1}}$, making the search cheaper than in the O1 search, which allows a more sensitive search at the same computing cost. 

In the future, we will combine RV measurements provided by our ongoing monitoring campaign of UVES/VLT observations with the 1999 and 2011 measurements to further refine the binary period and ephemeris and ensure a good ephemeris is available throughout the Advanced-LIGO era (see also G14).
Superior RV precision derived from future UVES data may also allow us to rule out low values of $\alpha$ (see Section~\ref{sec:masses}) by matching the observed residual RVs to the simulated residuals. 
This effort would directly improve the K-correction, and thereby improve the constraints on $K_2$ and $K_1$.
Additionally, we will seek to exploit the more complete phase coverage of the UVES/VLT data to compute Doppler tomograms for the VLT spectra, which may provide further insights into the value of $\mathrm{K_{disc}}$ $\simeq$ $K_1$.
The determination of $e$ will remain difficult due to our poor knowledge of both $q$ and $\alpha$.

\section*{Acknowledgements}

DS and TRM acknowledge support from STFC via grant ST/P000495/1. 
This project was supported in part by the Monash-Warwick Strategic Funding Initiative. D.K.G. acknowledges the support of the  Australian Research Council via the Future Fellowship scheme (project FT0991598).
We thank John Whelan and the rest of the LSC continuous-waves search group for useful discussions which helped shape this manuscript.
Based on observations made with ESO Telescopes at the La Silla Paranal Observatory under programme ID 087.D-0278.
 
 
 
\bibliographystyle{mnras}
\bibliography{LW_reflist}
 
 
 
 
\appendix
\section{Irradiation modelling}
%
Using the \emph{lprofile} code (within the $\textsc{lcurve}$ package) provided by T. Marsh (e.g. \citealt{2009A&A...507..929S}; \citealt{2010MNRAS.402.2591P}), we started by generating phase-resolved synthetic line profiles from an irradiated donor
for all possible $\alpha$-values and different mass ratios between the newly determined range of $q$ of 0.22--0.76.
In order to derive the most appropriate $V$$\mathrm{sin\emph{i}_{cor}}$ factors,
limb darkening effects were not included in the simulations for the optically thin regime examined here.
Other key input parameters representing the instrumental setup (e.g., the exposure length and the FWHM spectral resolution) were fixed at values that match our UVES/VLT observations.
Next, we fitted a 1D Gaussian profile to the model data sets and measured the widths of the emission components as a function of orbital phase, which always exhibit a single peak at phase 0.5
for all possible combinations of $\alpha$ and $q$.
Therefore, we take the width measurement at phase 0.5 (after correction for instrumental broadening)
as the estimate for the observed $\emph{V}\mathrm{sin}\emph{i}_\mathrm{{em}}$ that leads to the largest correction factor (i.e. smallest deviation).
Since both $q$ and true $K_2$ are known,
the true $V$sin$\emph{i}$ can be calculated using equation~(\ref{eq:Vsini}) and hence the ratio of $\frac{\emph{V}\mathrm{sin}\emph{i}_\mathrm{{em}}}{\emph{V}\mathrm{sin}\emph{i}}$ can be determined for each irradiation model.

Meanwhile, by establishing synthetic RV curves (through 1D Gaussian fits to the line profiles), 
it can clearly be seen that the RVs of the donor emission may deviate significantly from a sinusoid at high mass ratios or low disc opening angles
(more details will be discussed later in Section~\ref{sec:aE}).
We reconstructed synthetic Doppler maps to extract $\mathrm{K_{em}}$ (by computing the centroid of the donor emission spot; see Section~\ref{sec:MCD}), from which we could independently estimate $\mathrm{K_{cor}}$ = $\frac{\mathrm{K_{em}}}{K_2}$, to offer a comparison to MCM05.
 
\begin{figure}
        \includegraphics[width=\columnwidth]{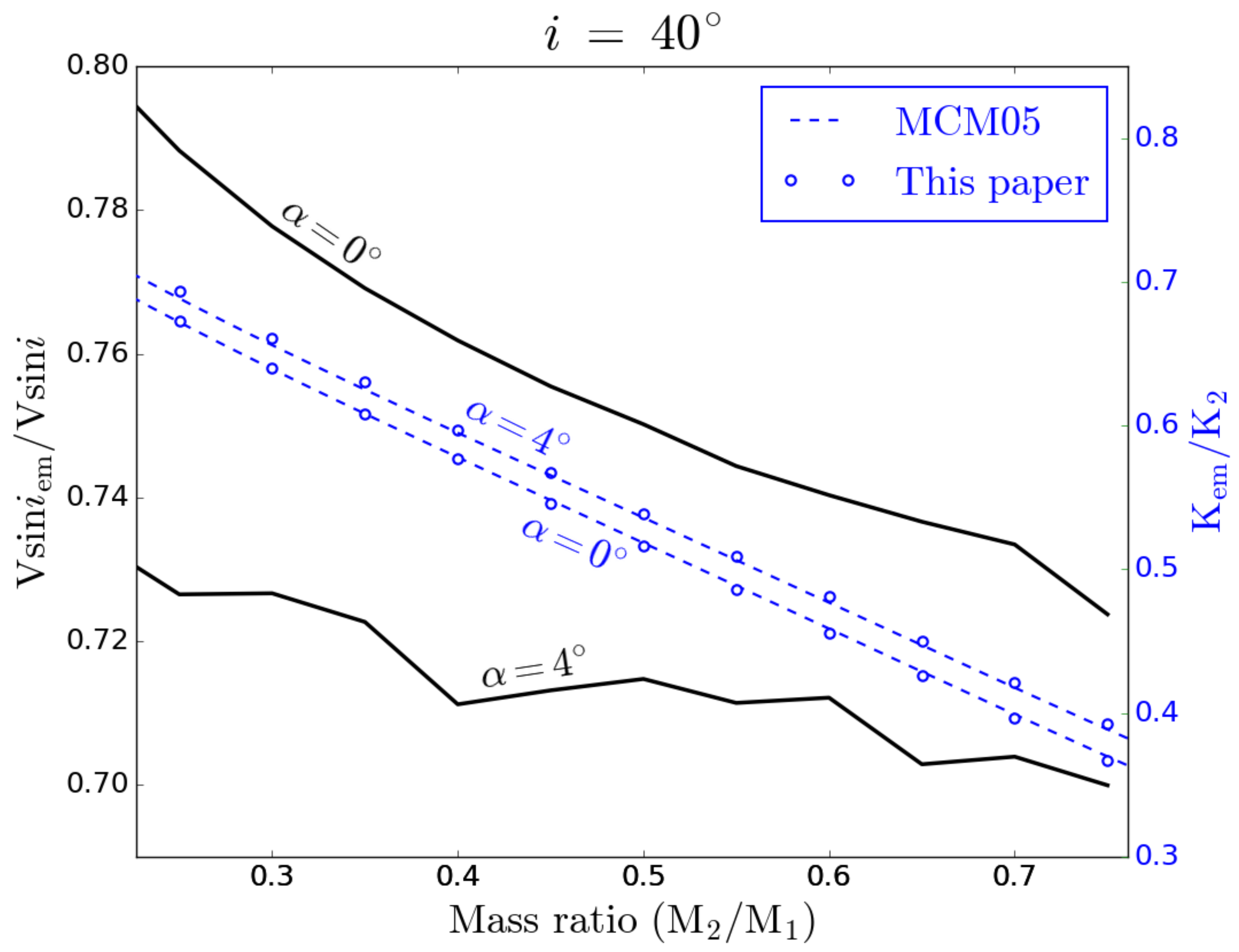}
    \caption{Plot of $V$$\mathrm{sin\emph{i}}$-- (black) and K-correction (blue) determined from our simulations (for the low inclination case of $i$ = $40^{\circ}$) as a function of $q$ for $\alpha$ = $0^{\circ}$ and $4^{\circ}$, with the fourth-order polynomial models provided in MCM05 overplotted (as blue dashed lines) for comparison. The minimum $V$$\mathrm{sin\emph{i}_{cor}}$-correction ($\alpha$ = $0^{\circ}$) should be applied to yield the most stringent lower limits of the system parameters ($q$, $K_1$ and $K_2$) of Sco X-1.}
    \label{fig:KcorVcor}
\end{figure} 

Fig.~\ref{fig:KcorVcor} presents example results of our simulations of $V$$\mathrm{sin\emph{i}_{cor}}$ (black) and $\mathrm{K_{cor}}$ (blue) versus $q$ (for the cases of $\alpha$ = $0^{\circ}$ and $4^{\circ}$),
with the K-correction of MCM05 produced by a different binary code overplotted (as blue dashed lines) for comparison. One can see that both factors are always significantly below unity, and this can be exploited to derive firmer limits.
The K-correction decreases smoothly with both $\alpha$ and $q$, and has already been approximated by a set of fourth-order polynomials by MCM05 for $\alpha$ from $0^{\circ}$ to $18^{\circ}$ in steps of $2^{\circ}$.
For small disc opening angles ($\alpha$ $\leq$ $4^{\circ}$), our solutions for $\mathrm{K_{cor}}$ coincide almost exactly with the numerical models of MCM05 (as shown in Fig.~\ref{fig:KcorVcor}).
The results were also in good agreement (within 3.5 per cent) for $\alpha$ $>$ $4^{\circ}$.
The $V$$\mathrm{sin\emph{i}}$-correction factor, on the other hand, tends to decrease as $\alpha$ increases; yet at any given $\alpha$-angle the estimated value of $V$$\mathrm{sin\emph{i}_{cor}}$ does not vary smoothly with $q$, fluctuating increasingly randomly with increasing $\alpha$.

\section*{Supporting information}
Supplementary data are available at \textit{MNRAS} online.

\bsp    
\label{lastpage}
\end{document}